\documentclass[letterpaper, 10 pt, conference]{ieeeconf}  

\IEEEoverridecommandlockouts                              

\usepackage{amsmath} 
\usepackage{amssymb}  

\usepackage{amsthm}
\usepackage{booktabs}
\usepackage[ruled,vlined]{algorithm2e}
\SetAlgorithmName{Procedure}{algorithm}{List of Procedures}

\usepackage{tikz}
\usetikzlibrary{shapes,arrows,calc,fit}
\tikzset{
        block/.style = {draw, rectangle,
            minimum height=1cm,
            minimum width=2cm},
        input/.style = {coordinate,node distance=1cm},
        output/.style = {coordinate,node distance=4cm},
        arrow/.style={draw, -latex,node distance=2cm},
        pinstyle/.style = {pin edge={latex-, black,node distance=2cm}},
        sum/.style = {draw, circle, node distance=1cm},
    }
\usepackage{pgfplots} 
\pgfplotsset{compat=newest} 
\pgfplotsset{plot coordinates/math parser=false} 
\newlength\figureheight 
\newlength\figurewidth 
\pgfplotsset{
    every axis plot post/.style={
        line join=round
    }
}
\usetikzlibrary{external} 

\newlength\defcolwidth

\newcommand{\includetikz}[1]{%
    \tikzsetnextfilename{#1}%
    \includegraphics[scale=1]{figures/#1.pdf}%
}

\usepackage{tikz}
\usepackage{tikzscale}

\definecolor{myorange}{cmyk}{0,0.35,0.85,0} 
\definecolor{mypurple}{cmyk}{0.5,1,0,0} 

\definecolor{matblue1}{rgb}{0,0.4470,0.7410}
\definecolor{matred1}{rgb}{0.85,0.325,0.098}
\definecolor{matyel1}{rgb}{0.9290, 0.6940, 0.1250}
\definecolor{matpur1}{rgb}{0.4940, 0.1840, 0.5560}
\definecolor{matgre1}{rgb}{0.4660, 0.6740, 0.1880}
\definecolor{matblue2}{rgb}{0.3010, 0.7450, 0.9330}
\definecolor{matred2}{rgb}{0.6350, 0.0780, 0.1840}
\definecolor{matgrey1}{rgb}{0.5, 0.6, 0.7}
\definecolor{matpink1}{rgb}{1, 0.07, 0.65}
\definecolor{matblue3}{rgb}{0.07, 0.62, 1}
\definecolor{gray09}{rgb}{0.9, 0.9, 0.9}

    \definecolor{mblue}{rgb}{0,0.447,0.741}
    \definecolor{mred}{rgb}{0.85,0.325,0.098}
    \definecolor{myellow}{rgb}{0.9290,0.6940,0.1250}
    \definecolor{mmagenta}{rgb}{1,0,1}
    \definecolor{mgreen}{rgb}{0.4460,0.6740,0.1880}
    \definecolor{mgrey}{rgb}{0.6,0.6,0.6}
    \definecolor{mpurple}{rgb}{0.4940, 0.1840, 0.5560}
    
    \tikzset{cross/.style={cross out, draw=black, minimum size=2*(#1-\pgflinewidth), inner sep=0pt, outer sep=0pt}, cross/.default={1pt}}
    
    \usetikzlibrary{shapes}

\newcommand{\blackdash}{\raisebox{2pt}{\tikz{\draw[-,black,dashed,line width = 0.9pt](0,0) -- (3mm,0);}}}

\newcommand{\blueline}{\raisebox{2pt}{\tikz{\draw[-,matblue1,solid,line width = 0.9pt](0,0) -- (3mm,0);}}}
\newcommand{\redline}{\raisebox{2pt}{\tikz{\draw[-,matred1,solid,line width = 0.9pt](0,0) -- (3mm,0);}}}

\newcommand{\purpleline}{\raisebox{2pt}{\tikz{\draw[-,mypurple,solid,line width = 0.9pt](0,0) -- (3mm,0);}}}

\newcommand{\yelline}{\raisebox{2pt}{\tikz{\draw[-,matyel1,solid,line width = 0.9pt](0,0) -- (3mm,0);}}}

\newcommand{\greenline}{\raisebox{2pt}{\tikz{\draw[-,matgre1,solid,line width = 0.9pt](0,0) -- (3mm,0);}}}
\newcommand{\bluelinet}{\raisebox{2pt}{\tikz{\draw[-,matblue2,solid,line width = 0.9pt](0,0) -- (3mm,0);}}}
\newcommand{\redlinet}{\raisebox{2pt}{\tikz{\draw[-,matred2,solid,line width = 0.9pt](0,0) -- (3mm,0);}}}

\newcommand{\greyarea}{\raisebox{0pt}{\tikz{\draw[-,gray09,solid,line width = 4pt](0,0) -- (3mm,0);}}}

\title{\LARGE \bf
Learning nonlinear feedforward: a Gaussian Process Approach Applied to a Printer with Friction 
}

\newtheorem{definition}{Definition}
\newtheorem{remark}{Remark}
\newtheorem{theorem}{Theorem}

\author{Max van Meer$^{1}$, Maurice Poot$^{1}$, Jim Portegies$^{2}$, Tom Oomen$^{1}$  
\thanks{$^{1}$Max van Meer (e-mail: m.v.meer@tue.nl), Maurice Poot and Tom Oomen are with the Control Systems Technology Group, Department of Mechanical Engineering, Eindhoven University of Technology, The Netherlands. Tom Oomen is also with the Delft Center for Systems and Control, Delft University of Technology, Delft, The Netherlands. This work is part of the research programme VIDI with project number 15698, which is (partly) financed by the Netherlands Organisation for Scientific Research (NWO). In addition, this research has received funding from the ECSEL Joint Undertaking under grant agreement 101007311 (IMOCO4.E). The Joint Undertaking receives support from the European Union’s Horizon 2020 research and innovation programme.}
\thanks{
$^{2}$Jim Portegies is with the Applied Analysis Group, Department of Mathematics and Computer Science, Eindhoven University of Technology, The Netherlands. }%
}
\let\labelindent\relax
    \usepackage{enumitem}
    
\begin{document}

\maketitle
\thispagestyle{empty}
\pagestyle{empty}

\begin{abstract}
Feedforward control is essential to achieving good tracking performance in positioning systems. The aim of this paper is to develop an identification strategy for inverse models of systems with nonlinear dynamics of unknown structure using input-output data, which directly delivers feedforward signals for a-priori unknown tasks. To this end, inverse systems are regarded as noncausal nonlinear finite impulse response (NFIR) systems, and modeled as a Gaussian Process with a stationary kernel function that imposes properties such as smoothness and periodicity. The approach is validated experimentally on a consumer printer with friction and shown to lead to improved tracking performance with respect to linear feedforward. 
\end{abstract}

\section{Introduction}
System identification in the presence of nonlinear dynamics of unknown structure is a challenging subject, because of the wide range of possible model descriptions that need to be considered \cite{Sjoberg1995}. For feedforward control, models of inverse systems are of particular interest, since the filtering of some reference by the inverse system yields the required control effort for that task. For a system $G$, an inverse model $G^{-1}$ can be obtained by either $(i)$ identification of $G$ and subsequent inversion, or $(ii)$ direct identification of $G^{-1}$. It is shown in \cite{Blanken2020} that the latter poses an advantage over the former, since properties such as stability, smoothness and finite preview or history of $G^{-1}$ can be enforced directly on the model. 

If the nonlinear structure of the inverse system is known to be representable by, e.g., a set of some polynomial basis functions, the model coefficients can be learned perfectly in an iterative fashion, see \cite{Boeren2015,VanDeWijdeven2010}. Such an approach allows for the generation of feedforward signals for any task, because any reference can be filtered through this inverse model to obtain the required control effort. 

Identification methods for inverse systems that rely on user-specified parametrization of the system through a limited number of nonlinear basis functions are not applicable when the nonlinear structure of a system is unknown. In such case, the structural model errors of the feedforward controller can lead to performance degradation \cite{Schoukens2019}.

\begin{figure}[!t]
\centering
\includetikz{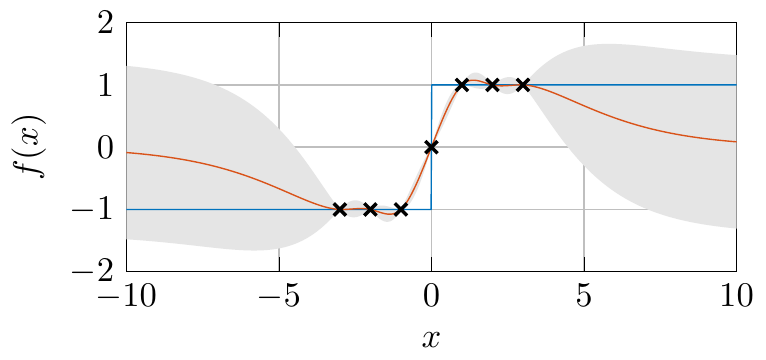}
\caption{An example function (\protect\blueline) modeled as a Gaussian Process. The posterior distribution, using a Matèrn$_{3/2}$ kernel, has an accurate mean (\protect\redline) close to observations (\protect$\boldsymbol{\times}$), and high standard deviation (\protect\greyarea) far from observations.}
\label{fig:posterior_example}
\end{figure}

For a certain class of basis functions, namely, the eigenfunctions of universal kernels, structural model errors of a modeled function $f:\mathbb{R}^n\rightarrow\mathbb{R}$ on a compact subset $\mathcal{X}\subset\mathbb{R}^n$ vanish as the number of basis functions grows to infinity \cite{Micchelli2006}. These kernels impose properties such as smoothness on the model in order to deal with the bias-variance trade-off. In Gaussian Process (GP) regression, these kernels pose a Gaussian prior on $f$, and predictions are made by extrapolating from measurements using Bayes' rule \cite{Rasmussen2004}, see Figure \ref{fig:posterior_example}. 

GPs have been shown in literature to be applicable to modeling dynamic systems with unknown nonlinear dynamics. Such modeling methods are widely available for Euler-Lagrange systems \cite{Beckers2019,Nguyen-Tuong2008,Nguyen-Tuong}, single-input systems that offer full state measurements \cite{Deisenroth2011}, and causal systems \cite{Pillonetto2014}.

Although these methods are capable of modeling dynamics with unknown nonlinear structure, the requirement of full state measurements is overly restrictive in some cases, and in the presence of nonlinear dynamics of unknown structure, the inclusion of a state observer is nontrivial. Moreover, for motion systems $G$, the inverse $G^{-1}$ is always noncausal, and for reasons that will become clear in Section \ref{sec:kernel}, nonlinear kernel-based identification methods of $G$ may not be directly applicable to identification of $G^{-1}$.

The aim of this paper is to develop a novel technique for grey-box modeling of noncausal nonlinear systems with unknown structure based on input-output data. In particular, inverse systems $G^{-1}$ are viewed as noncausal nonlinear finite impulse response (NFIR) systems and modeled as a Gaussian Process, in order to generate feedforward signals for a range of tasks that may be unknown a-priori. This paper consists of two main contributions:\begin{enumerate}[label={C\arabic*:}]
\item A procedure for GP-based feedforward control of systems with unknown nonlinear dynamics is proposed and described from a design perspective.
\item The method is validated experimentally on a printer with friction to show its improved tracking performance with respect to linear feedforward control.
\end{enumerate}

This paper is structured as follows. First, the problem description is given in Section \ref{sec:problem}. Subsequently, Gaussian Processes regression for NFIR systems is explained and design considerations for kernel selection and experimental design are given in Section \ref{sec:kimcon}. Afterwards, in Section \ref{sec:printer} the method is applied to an experimental setup. Finally, Section \ref{sec:conclusions} presents the conclusions and proposes some directions for future work.

\tikzstyle{block} = [draw, fill= white!20, rectangle, 
    minimum height=3em, minimum width=3em]
\tikzstyle{sum} = [draw, fill=white!20, circle, node distance=1cm]
\tikzstyle{input} = [coordinate]
\tikzstyle{output} = [coordinate]
\tikzstyle{pinstyle} = [pin edge={to-,thin,black}]
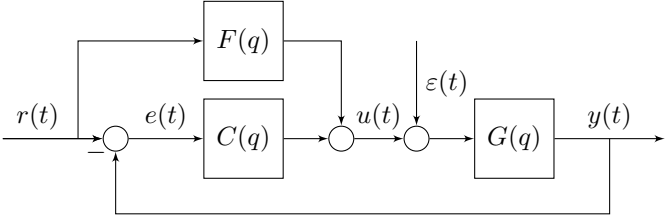
\begin{figure}[!t]
\centering
\begin{tikzpicture}[auto, node distance=2cm,>=latex']

            \node [input, name=input] {};

            \node [sum, right of= input, node distance=1.5cm] (sum) {};
            \node [block, right of= sum, node distance=1.7cm] (controller) {$C(q)$};
                        \node [sum,right of = controller,node distance=1.3cm] (summid) {};

                        \node [sum,right of = summid] (sum2) {};

            \node [block, right of= sum2,node distance=1.3cm] (plant) {$G(q)$};
            
            \node [output, right of= plant] (output) {};
            \node [block, above of=controller,node distance=1.3cm] (ff) {$F(q)$};
                        \node [input, name=disturbance,above of=sum2,node distance=1.3cm] {};

            \draw [draw,->] (input) -- node [pos=0.35]{} (sum);
            \draw [draw,->] (disturbance) -- node []{$\varepsilon(t)$} (sum2);

            \draw [draw,->] (input) -- node [pos=0.35]{$r(t)$} (sum);
            \draw [->] (sum) -- node [name=e] {$e(t)$} (controller);
            \draw [->] ($(sum) + (-0.5cm,0)$) |- node {} (ff);

            \draw [->] (controller) -- node {} (summid);
            \draw [->] (summid) -- node {$u(t)$} (sum2);
                        \draw [->] (sum2) -- node {} (plant);

                        \draw [->] (ff) -| node {} (summid);
            \draw [->] (plant) -- node [name=y] {$y(t)$}(output);
            \draw [->] (y) -- ++ (0,-1.3) -| node [pos=0.99] {$-$} (sum);
\end{tikzpicture}
\caption{Motion control architecture.}\label{fig:setting_closed}
\end{figure}
\section{Problem description}\label{sec:problem}
In this section, the problem description is given. First, the control architecture and goal are described. Subsequently, an assumption on the structure of the inverse system is explained and it is shown how this setting allows for task flexibility. 
\subsection{Setting and goal}\label{sec:setting}
Let $G(q)$ denote a discrete-time, nonlinear SISO system, where $q$ is the forward-shift operator such that $q^\tau a(t) := a(t+\tau)$, $\tau\in\mathbb{Z}$. The closed-loop motion control architecture in Figure \ref{fig:setting_closed} is considered, where $r(t)$ denotes a predefined reference to be tracked at time $t\in\mathbb{Z}$, $y(t)$ is the system output, $e(t)$ is the tracking error and $\varepsilon(t)\sim\mathcal{N}(0,\sigma_n^2)$ is a disturbance. \\
It is assumed that a stabilizing feedback controller $C(q)$ is available, as well as some prior feedforward controller $F(q)$. Perfect tracking is achieved for $F(q)=G^{-1}(q)$, since then \begin{equation}
\begin{aligned}
e(t)&=r(t)-y(t)\\
&=r(t)-G(q) \left(u(t)+\varepsilon(t)\right)\\
&=r(t)-G(q) \left(G^{-1}(q) r(t)+ \varepsilon(t)\right)\\
\mathbb{E}[e(t)]&=0.
\end{aligned}
\end{equation} 

The goal in this paper is to obtain $\hat{F}(q)\approx G^{-1}(q)$ from data such that the 2-norm of the tracking error ($\|e\|_2$) is reduced. Moreover, task flexibility must be allowed.\begin{definition}[Task flexibility]
Task flexibility refers to the notion that the designed feedforward controller $\hat{F}(q)$ must achieve improved tracking performance for a range of different tasks that may not all be known prior to the generation of $\hat{F}(q)$.
\end{definition}
The next section explains an additional assumption on the structure on $G^{-1}(q)$, before the problem formulation is formalized in Section \ref{sec:problemform}.

\subsection{Noncausal nonlinear finite impulse response systems}
In order to find a feedforward controller $\hat{F}(q)\approx G^{-1}(q)$, we first impose some structure on $G^{-1}(q)$. First, note that \emph{linear} sampled systems $P(z)$ of continuous-time systems $P(s)$ often contain zeros outside the unit disc, depending on the sample time and the relative degree of $P(s)$, see \cite{Astrom1984}. Consequently, $P^{-1}(z)$ may contain poles outside the unit disc. Whereas systems with poles outside the unit disc are typically viewed as causal and unstable, they may also be viewed as noncausal and stable.

\begin{theorem}[Non-causal exact inversion for NMP systems \cite{Blanken2020}]
Let system $P(z)$ be given such that $P^{-1}(z) \in$ $\mathcal{R} \mathcal{L}_{2}(\mathbb{T})$, i.e., the set of real, rational, discrete-time systems without poles on the unit disc $\mathbb{T}:=\{z \in \mathbb{C}:|z|=1\}$. Then, there exists a non-causal sequence $\theta^{} \in$ $\ell_{1}(\mathbb{Z})$ such that, for any signal $r(t) \in \ell_{2}(\mathbb{Z})$, the signal \begin{equation}\label{eq:iir}
u_{}(t)=\sum_{\tau=-\infty}^{\infty} \theta_{\tau}^{} r(t-\tau) \in \ell_{2}(\mathbb{Z})
\end{equation}
leads to exact inversion $y_{}(t)=P(q) u_{}(t)=r(t)$.\end{theorem}

In the absence of infinite preview or history, a non-causal FIR system parametrization may be used as a finite-dimensional approximation of \eqref{eq:iir}: \begin{equation}\label{eq:fir}
\begin{aligned}
u_{}(t)&\approx\sum_{\tau=-n_{ac}}^{n_c} \theta_{\tau}^{} r(t-\tau)\\
&\approx \theta^\top \mathbf{r}_t =: f_{\text{lin}}(\mathbf{r}_t),
\end{aligned} 
\end{equation}
with $
\mathbf{r}_{t}:=\left[r\left(t+n_{a c}\right), \ldots, r\left(t-n_{c}\right)\right]^{\top}\in\mathcal{Y}$, where $\mathcal{Y}\subset\mathbb{R}^
{n_{\theta}}$ is compact with $n_{\theta}=n_c+n_{ac}+1$, for reasons described in Section \ref{sec:structure}. \\

In this paper, this idea is extended to nonlinear systems by assuming that the nonlinear system $G^{-1}(q)$ can be represented with a non-causal nonlinear finite impulse response (NFIR) parametrization, such that the control effort $u(t)$ required to realize output sequence $
\mathbf{y}_{t}:=\left[y\left(t+n_{a c}\right), \ldots, y\left(t-n_{c}\right)\right]^{\top}\in\mathcal{Y}$ can be  written as \begin{equation}\label{eq:nfir}
u(t)=f\left(\mathbf{y}_{t}\right)+\varepsilon(t),
\end{equation}
where $f:\mathcal{Y}\rightarrow \mathbb{R}$ is an alternative form of $G^{-1}(q): \mathbb{R}\rightarrow\mathbb{R}$, in which $q$ has been eliminated. 
\subsection{Problem formulation}\label{sec:problemform}
In this paper, the aim is to model the function $f$ in \eqref{eq:nfir} describing $G^{-1}$ from a dataset $\mathcal{D}=\{u(t),y(t)\}_{t=1}^M$, such that $f(\mathbf{r}_t)$ yields the control effort (or feedforward signal) $u(t)$ that realizes reference sequence $
\mathbf{r}_{t}:=\left[r\left(t+n_{a c}\right), \ldots, r\left(t-n_{c}\right)\right]^{\top}\in\mathcal{Y}$. Note that by learning $f$, feedforward samples $u(t)$ can be computed for any $\mathbf{r}_{t}$, i.e., task flexibility is allowed. 
\section{Nonlinear feedforward using Gaussian Process Regression}\label{sec:kimcon}
In this section, Gaussian Process regression is employed to generate feedforward signals for nonlinear systems. First, it is explained how NFIR systems can be represented by Gaussian Processes. Second, the choice of kernel function is addressed from a design perspective. Finally, it is shown how a data-set can be obtained that allows for task flexibility.

\subsection{Gaussian Process models of NFIR systems}
Next, $f$ in \eqref{eq:nfir} is regarded as a {Gaussian Process} (GP), to learn it from data. 
\begin{definition}[Gaussian Process \cite{Rasmussen2004}]
A Gaussian Process is defined as an indexed family of random variables $g(\mathbf{x})\in\mathbb{R}$ with $\mathbf{x}\in\mathcal{X}$, any finite number of which have a joint Gaussian distribution.
\end{definition}  

Hence, if $f(\mathbf{y}_t)$ is a GP, then there exists a joint distribution between \emph{observations} of $u(t)=f(\mathbf{y}_t)+\varepsilon(t)$ and \emph{unknown feedforward samples} $u(t)=f(\mathbf{r}_t)$. It is shown next how GPs can be employed to make predictions of these feedforward samples, based on observations of $u$ and $y$.  \\

Let the covariance between any two control effort values required for output sequences $\mathbf{y}_t$ be defined by a kernel function\begin{equation}\label{eq:kernel}
k(\mathbf{y}_{t_1},\mathbf{y}_{t_2}) := \operatorname{cov}(f(\mathbf{y}_{t_1}),f(\mathbf{y}_{t_2})).
\end{equation}
This kernel function imposes properties such as smoothness or periodicity on $f$, as explained in detail in Section \ref{sec:kernel}. Given a dataset $\mathcal{D}=\left\{Y,\mathbf{u}\right\}$, with\begin{equation}\label{eq:data}
\begin{aligned}
Y&=\left[\mathbf{y}_{1}, \ldots, \mathbf{y}_{M}\right]^{\top}, \\
\mathbf{u}=\mathbf{f}(Y)&=[u(1), \ldots, u(M)]^{\top},
\end{aligned}
\end{equation}
and assuming a zero-mean prior on $f$, we can define the joint distribution \begin{equation}\label{eq:joint}
\left[\begin{array}{c}
\mathbf{f}(Y) \\
\mathbf{f}(R)
\end{array}\right] \sim \mathcal{N}\left(\mathbf{0},\left[\begin{array}{cc}
K(Y,Y)+\sigma_{n}^{2} & K\left(Y,R\right) \\
K\left(R,Y\right) & K\left(R,R\right)
\end{array}\right]\right),
\end{equation}
where $R=\left[\mathbf{r}_{1}, \ldots, \mathbf{r}_{N}\right]^{\top}$ is a toeplitz matrix constructed from $r \in \mathbb{R}^{N}$, with $\mathbf{r}_{t}:=\left[r\left(t+n_{a c}\right), \ldots, r\left(t-n_{c}\right)\right]^{\top}$. The covariance matrices $K$ in \eqref{eq:joint} are constructed by evaluating $k$ for each element. \\

To compute the feedforward signal $\mathbf{f}(R)$ for reference $R$, the joint distribution in \eqref{eq:joint} is conditioned on the observations using Bayes' rule. The resulting {posterior distribution} $p(\mathbf{f}(R)\mid Y,\mathbf{u},R)$ is a Gaussian with mean and variance
\begin{equation}\label{eq:posterior}
\begin{aligned}
\mathbb{E}\left[\mathbf{f}(R)\right] &=K(R,Y)\left[K(Y,Y)+\sigma_{n}^{2} I\right]^{-1} \mathbf{u},\\
\operatorname{cov}(\mathbf{f}(R)) &=K(R, R) \\
&-K(R, Y)^{\top}\left[K(Y, Y)+\sigma_{n}^{2} I\right]^{-1} K(Y, R).
\end{aligned}
\end{equation}
This yields an analytic expression for the expected feedforward signal required for reference $r$, as well as explicit uncertainty bounds on the feedforward signal.

\subsection{Model structure}\label{sec:structure}
Whether the realization $\mathbb{E}[\mathbf{f}(R)]$ is representative of the real function $f$ is dependent on the choice of $k$ in \eqref{eq:kernel}. This can be seen easily by rewriting the posterior mean \eqref{eq:posterior} in scalar form as \begin{equation}\label{eq:posterior_alpha}
u(t) = \mathbb{E}\left[f(\mathbf{r}_t)\right] = \sum_{i=1}^M \alpha_i k(\mathbf{y}_i,\mathbf{r}_t),
\end{equation}
with $\boldsymbol{\alpha}=(K(Y,Y)+\sigma_n^2 I)^{-1} \mathbf{u}$. Naturally, if $k$ were to be chosen inappropriately, e.g., linear in $\mathbf{r}_t$, as done for linear systems in \cite{Blanken2020}, no nonlinear function $f$ could be represented by the posterior mean. A particularly widely applicable type of kernel is the \emph{universal kernel}. 
\begin{definition}[Universal kernel \cite{Micchelli2006}]\label{def:universal}
Given any compact subset $\mathcal{Z}$ of $\mathcal{Y}$, let $C(\mathcal{Z})$ be the space of all continuous complex-valued functions from $\mathcal{Z}$ to $\mathbb{C}$ with maximum norm $\|.\|_\mathcal{Z}$. More over, let $K(\mathcal{Z})$ denote all functions in $C(\mathcal{Z})$ which are uniform limits of functions of the form \eqref{eq:posterior_alpha} where $\{\mathbf{y}_i : i\in\mathbb{N}_n\}\subseteq \mathcal{Z}$. If for any such $\mathcal{Z}$, any positive number $\epsilon$ and any function $f\in C(\mathcal{Z})$, there is a function $g\in K(\mathcal{Z})$ such that $\|f-g\|_\mathcal{Z}\leq \epsilon$, then $k$ is a universal kernel. 
\end{definition} 

Hence, universal kernels are structurally capable of representing any continuous $f$ through \eqref{eq:posterior_alpha}, even if the nonlinear structure of $f$, or equivalently, the non-causal NFIR system $G^{-1}(q)$, is unknown. 

In practice, some information on $f$ may be available, such as smoothness or periodicity. This information can be exploited through \emph{stationary} kernel functions, as explained in more detail in the next section. 
\begin{definition}[Stationary kernel function \cite{Rasmussen2004}]
A kernel $k$ is stationary if $k(\mathbf{y}_{t_1},\mathbf{y}_{t_2})=k(\mathbf{y}_{t_1}-\mathbf{y}_{t_2})$.
\end{definition}
This choice of kernel leads to a data-dependent model \eqref{eq:posterior_alpha} in which control effort values $u(t) = f(\mathbf{r}_t)$ are inferred from observations of $u(t)$ corresponding to similar sequences $\mathbf{y}_t\approx \mathbf{r}_t$, i.e., similar paths in $\mathcal{Y}$ require similar control effort values. The next section lists some examples of stationary kernels from a design perspective.

\subsection{Kernel selection}\label{sec:kernel}
\begin{figure}[!t]
\centering
\includetikz{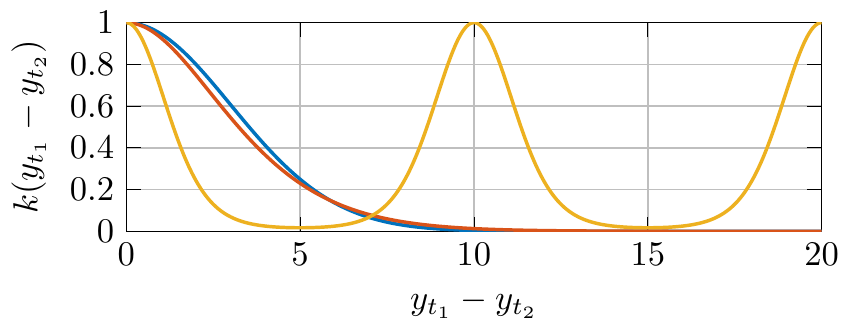}
\caption{Comparison of the covariance of a squared-exponential kernel (\protect\blueline), a Matèrn$_{3/2}$ kernel (\protect\redline) and a periodic kernel (\protect\yelline) along one dimension of $\mathbf{y}_{t_1}-\mathbf{y}_{t_2}$. Similar output sequences $\mathbf{y}_t$ require similar control effort values $u(t)$, as can be seen from the high covariance near $y_{t_1}=y_{t_2}$.}
\label{fig:kernel_compare}
\end{figure}

This section lists some example kernels, distinguishing between two cases: $(i)$ $f$ is non-periodic, in which a universal kernel is chosen based on smoothness of $f$, and $(ii)$ $f$ is known to be periodic, in which a non-universal kernel is used to exploit the periodic structure. The kernels are compared visually in Figure \ref{fig:kernel_compare}. 

\subsubsection{Smooth dynamics}
In this paper, dynamics are considered \emph{smooth} if $f$ is infinitely differentiable. A kernel that imposes smoothness on $f$ by being infinitely differentiable is the \emph{squared-exponential kernel}: \begin{equation}\label{eq:se}
k\left(\mathbf{y}_{t_1}, \mathbf{y}_{t_2}\right)=\sigma_{f}^{2} \exp \left(-\frac{1}{2} \rho\right),
\end{equation}
where \begin{equation}
\rho=\left(\mathbf{y}_{t_1}-\mathbf{y}_{t_2}\right)^{\top} \Lambda^{-1}\left(\mathbf{y}_{t_1}-\mathbf{y}_{t_2}\right),
\end{equation}
in which hyper-parameters $\sigma_{f}^{2}=\operatorname{Var}(f(\mathbf{y}_t))$ relate to the maximum magnitude of the control effort and $\Lambda=\operatorname{diag}\left(\left[\ell_{1}, \ldots, \ell_{n_{\theta}}\right]\right)$ contains kernel length-scales $\ell_i$. This kernel is universal, see \cite{Sriperumbudur2011}.
\subsubsection{Non-smooth dynamics}
Alternatively, $f$ may be finitely differentiable, e.g., as a result of static friction. In this case, the Matèrn$_{3/2}$ kernel function can be used, which is able to represent non-smooth functions since it is only 1-time mean square differentiable \cite[Section 4.2]{Rasmussen2004}: 
\begin{equation}\label{eq:matern}
k\left(\mathbf{y}_{t_1}, \mathbf{y}_{t_2}\right)=\sigma_{f}^{2}(1+\sqrt{3} \rho) \exp (-\sqrt{3} \rho).
\end{equation}
This is a universal kernel as well, see \cite{Sriperumbudur2011}.
\subsubsection{Periodic dynamics}
If $f(\mathbf{y}_t)$ is periodic in elements $i$ of $\mathbf{y}_t$ with period $p_i$, a periodic kernel may be chosen \cite{Mackay2004}:  
\begin{equation}
\begin{aligned}
k\left(\mathbf{y}_{t_1}, \mathbf{y}_{t_2}\right)&=\sigma_{f}^{2} \exp \left[-\frac{1}{2}\right. \cdot \\
&\left.\sum_{i=1}^{n_{\theta}}\left(\frac{\sin \left(\frac{\pi}{p_{i}}\left(y_{t_1,i}-y_{t_2,i}\right)\right)}{\ell_{i}}\right)^{2}\right].
\end{aligned}
\end{equation}
This kernel is not universal, since it does not represent non-periodic functions, see \cite[Proposition 10]{Sriperumbudur2008}. If $f(\mathbf{y}_t)$ is periodic in some elements of $\mathbf{y}_t$ but not in others (e.g., dynamics that are periodic in position but not in velocity), this periodic kernel can be added to a different kernel, e.g., \eqref{eq:se} or \eqref{eq:matern}, see \cite[Section 4.2.4]{Rasmussen2004}. \\

\subsection{Hyper-parameter tuning}\label{sec:marginal}
The kernel hyper-parameters $\Theta=\{\sigma_n$ $\sigma_f$, $\ell_i\}$ can be optimized automatically by maximization of the log-marginal likelihood \cite[Section 5.4]{Rasmussen2004}. This is the probability of the data given the model, defined as \begin{equation}\label{eq:marginal}
\log p(\mathbf{u} \mid Y, {\Theta})=-\frac{1}{2} \mathbf{u}^{\top} K_{n}^{-1} \mathbf{u}-\frac{1}{2} \log \left|K_{n}\right|-\frac{M}{2} \log 2 \pi,
\end{equation}
with $K_{n}=K+\sigma_{n}^{2} I$ and $\left|K_{n}\right|:=\operatorname{det}\left(K_{n}\right)$. A local maximizer $\hat{\Theta}$ of the non-convex reward \eqref{eq:marginal} is obtained through active-set optimization \cite[Section 7.4]{Design2013}.
\subsection{Convergence}\label{sec:convergence}
As the density $\beta$ of the data tends to infinity, expressed in number of observations ($\mathbf{y}_t$,$u(t)$) per unit of $\mathbf{y}_t$-space (e.g., [m$^{n_{\theta}}$]), the posterior mean of a GP with a universal stationary kernel converges to the true function \cite[Section 7.1]{Rasmussen2004}. In the setting described in Section \ref{sec:setting}, it not possible to obtain a data-set of observations $\mathbf{y}_t$ distributed uniformly over $\mathcal{Y}$, since for sampled motion systems, each observation exhibits $y(t+1)\approx y(t)$. Still, given that the observations of $\mathbf{y}_t$ are close to $\mathbf{r}_t$ in $\mathbb{R}^{n_\theta}$, it follows from \cite[Section 1]{Sollich2005} that \begin{equation}\label{eq:convergence}
\lim_{M\rightarrow\infty} \sum_{i=1}^M \alpha_i k(\mathbf{y}_i,\mathbf{r}_t) = f(\mathbf{r}_t),
\end{equation}
if $k$ is a stationary universal kernel with sufficiently small length-scales $\ell_i$.

The next section describes how observations of $\mathbf{y}_t\approx\mathbf{r}_t$ can be obtained such that the posterior approximately converges according to \eqref{eq:convergence}.

\begin{figure}[!t]
\centering
\includetikz{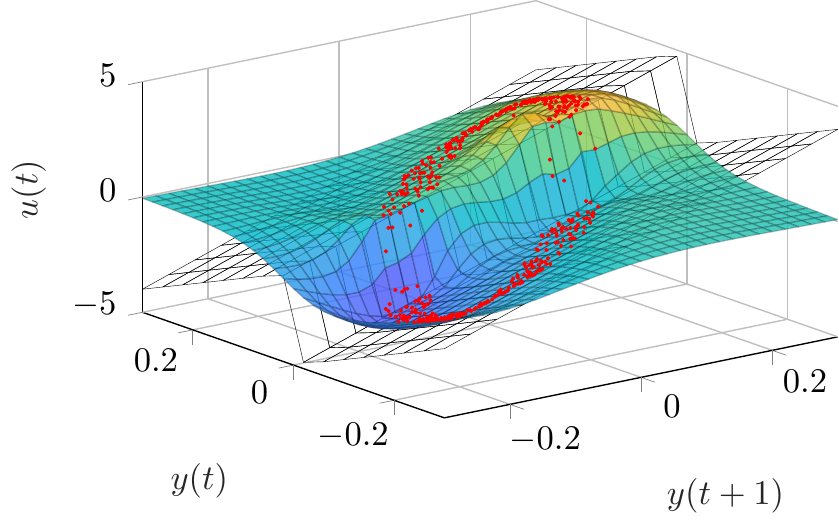}
\caption{Nonlinear example system of a mass with friction described by $u(t)=\frac{m}{T_s}(y(t+1)-y(t))+F_c \operatorname{sign}(y(t))$ (transparent mesh), modeled as a GP. The posterior mean (coloured mesh) is computed using a dataset containing observations $\left(\mathbf{y}_t,\ u(t)\right)$ (red dots) extracted from some periodic trajectories. The posterior mean is only accurate near the data, i.e., only feedforward signals for similar periodic references can be predicted accurately.}
\label{fig:sign}
\end{figure}
\subsection{Experiment design for task flexibility}\label{sec:data}
By virtue of the system parametrization in terms of a stationary covariance function, predictions of $f(\mathbf{r}_t)$ are inferred from observations $f(\mathbf{y}_t)$ of similar output sequences $\mathbf{y}_t\approx \mathbf{r}_t$. Hence, this method allows for task flexibility in so far that feedforward signals for different reference sequences $\mathbf{r}_t$ can be predicted accurately if similar sequences are present in $\mathcal{D}$. \\

To collect a dataset that allows for the generation of feedforward signals for flexible tasks, it is proposed to design several signals $\tilde{\mathbf{r}}_j\in\tilde{\mathcal{R}}$ to be used as a reference in closed-loop experiments to obtain $\mathcal{D}$, see Figure \ref{fig:setting_closed}.
\begin{remark}
In closed-loop, the noise $\varepsilon(t)$ introduces correlation between $u$ and $y$ through feedback, leading to bias in the estimate of $G^{-1}$. This bias could be reduced by repeating each experiment such as to observe a different realization of the noise, see \cite{Boeren2015}.
\end{remark}
The proposed method is summarized in Procedure \ref{proc:kimcon}.
\begin{algorithm}[]
\SetAlgoLined
\KwResult{Feedforward signal $\mathbf{u}_{\text{ff},i}$ for every reference ${r}_i\in\mathcal{R}$.}
\textbf{Input:} references $\mathcal{R}$, signals $\tilde{\mathcal{R}}$, initial feedforward controller $F(q)$, feedback controller $C(q)$.\\
 Define a suitable kernel function $k$\;
 Define $n_c,\ n_{ac}$\;
 \For{each signal $\tilde{{r}}_j$ in $\tilde{\mathcal{R}}$}{
  Conduct a closed-loop experiment with ${r}_j$ to obtain observations $\mathbf{y}_t$ and $u(t)$\;
  }
  Format the data to obtain $\mathcal{D}=\{Y,\mathbf{u}\}$ using \eqref{eq:data}\;
  Optimize for the kernel hyper-parameters, see Section \ref{sec:marginal}\;
  \For{each reference ${r}_i$ in $\mathcal{R}$}{
  Compute the posterior mean $\mathbf{u}_{\text{ff},i}=\mathbb{E}[\mathbf{f}(R)]$ using \eqref{eq:posterior}.
 }
 \caption{GP-based feedforward for nonlinear systems}\label{proc:kimcon}
\end{algorithm}
It is stressed that this procedure contains no recursion and hence there can be no issues with stability. Whether the posterior mean $\mathbb{E}[f]$ converges to the true $f$ (see Section \ref{sec:convergence}) now solely depends on the chosen $C(q),\ F(q)$ and $\tilde{\mathcal{R}}$. The next section shows that even with simple choices of $C(q),\ F(q)$ and $\tilde{\mathcal{R}}$ the procedure can lead to improved tracking performance. 

\section{Experimental validation}\label{sec:printer}
The proposed GP-based method for nonlinear feedforward control is applied to a desktop printer with friction, to demonstrate its ability to learn nonlinear dynamics for flexible tasks.
\subsection{Setting and goal}
\begin{figure}[!t]
\centering
\setlength{\fboxsep}{0pt}%
\fbox{\includegraphics[width=\figurewidth]{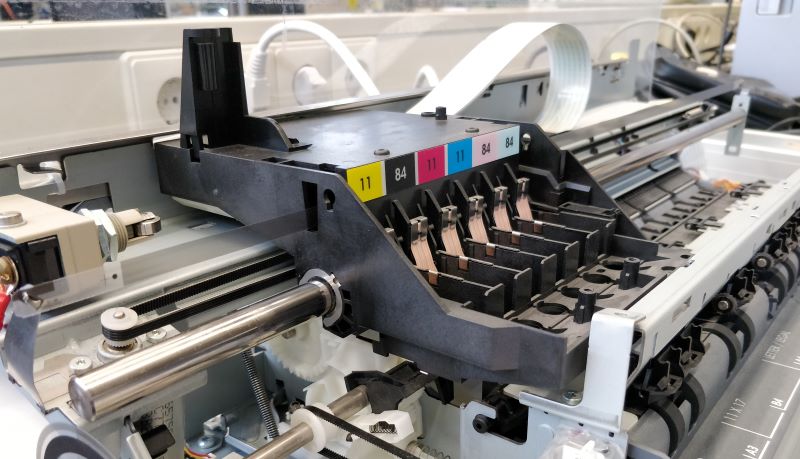}}
\caption{The desktop printer used for experimental validation. The motor on the bottom left drives the pulley, which is connected to the print-head by the toothed belt.}
\label{fig:printer}
\end{figure} 
The experimental set-up consists of a current-controlled A3 printer subject to static friction, depicted in Figure \ref{fig:printer}, connected to a computer running Simulink with a sample frequency of 1 kHz. A feedback controller $C(q)$ is available, as well as an initial feedforward controller $F(q)$ of the form \begin{equation}
F(q) = 2.8531 \frac{q-1}{q T_s} + 0.083\left(\frac{q-1}{q T_s}\right)^2,
\end{equation}
i.e., a controller with velocity and acceleration as basis. The goal is to track two third-order references ${r}_1$ and ${r}_2=1.05{r}_1$ of length $N=4501$ samples, depicted in Figure \ref{fig:ref}, using the closed-loop scheme of Figure \ref{fig:setting_closed}. To this end, the GP-based approach described in Section \ref{sec:kimcon} is applied to learn a new feedforward controller $\hat{F}(q)$.
\subsection{Approach}
Following Procedure \ref{proc:kimcon}, a kernel function is defined first. The Matèrn$_{3/2}$ kernel \eqref{eq:matern} is chosen, because it is expected that static friction leads to a non-smooth function $f(\mathbf{y}_t)$. The number of samples history and preview are defined as $n_c=20$ and $n_{ac}=40$ respectively. 

\begin{figure}[!t]
\centering
\includetikz{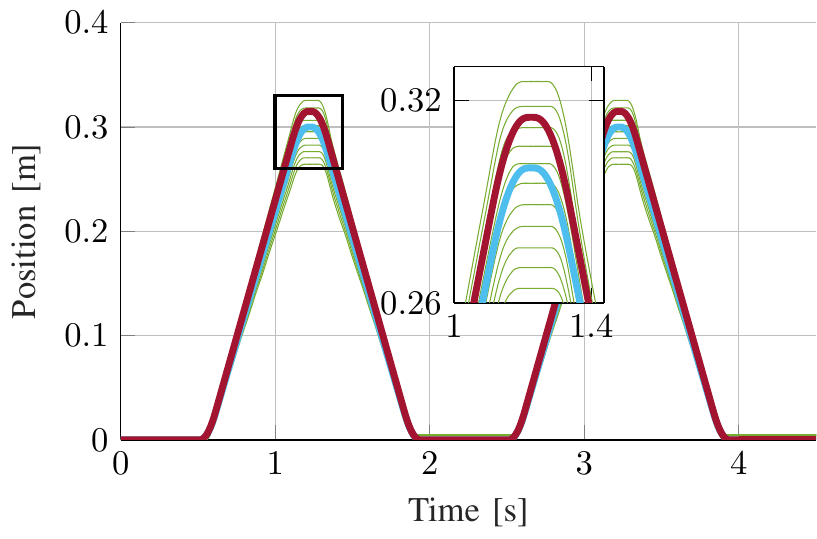}
\caption{References ${r}_1$ (\protect\bluelinet) and ${r}_2$ (\protect\redlinet) and observed outputs ${y}_i$ (\protect\greenline) from 11 closed-loop experiments using signals $\tilde{{r}}_i\in\tilde{\mathcal{R}}$ as a reference. Feedforward signals can be generated for different references $r$ with this data-set, demonstrating task flexibility.}
\label{fig:ref}
\end{figure}

Subsequently, 11 signals $\tilde{{r}}_{j}$ are designed as scaled variations of ${r}_1$, such that \begin{equation}
\tilde{\mathcal{R}}: \tilde{{r}}_{j}=a_{j} {r}_{1}, a_j \in[0.90,0.92, \ldots, 1.10].
\end{equation}
Note that this leads to $\tilde{{r}}_6={r}_1\in\mathcal{R}$, i.e., the first final reference ${r}_1$ is used during training, unlike $r_2$. \\
Each signal $\tilde{r}_i\in\tilde{\mathcal{R}}$ is used as a reference for one closed-loop experiment using $C(q)$ and $F(q)$, the result of which is shown in Figure \ref{fig:ref}. The observations of $u(t)$ and $y(t)$ are used to form $\mathcal{D}$ using \eqref{eq:data}, assuming $u(\tau)=y(\tau)=0\ \forall\tau < 0,\ \tau>N$. For computational reasons, the size of the dataset is reduced by only taking every $30$ rows of $Y$ and $\mathbf{u}$ into account to form $\mathcal{D}$. This leads to a dataset of $M=2970$ observations of $\mathbf{y}_t$ and $u(t)$. The kernel hyper-parameters are optimized by maximization of the log-marginal likelihood (see Section \ref{sec:marginal}) and the feedforward signals $\mathbf{u}_{\text{ff},1}$, $\mathbf{u}_{\text{ff},2}$ are computed with \eqref{eq:posterior}. 

\begin{figure}[!t]
\centering
\includetikz{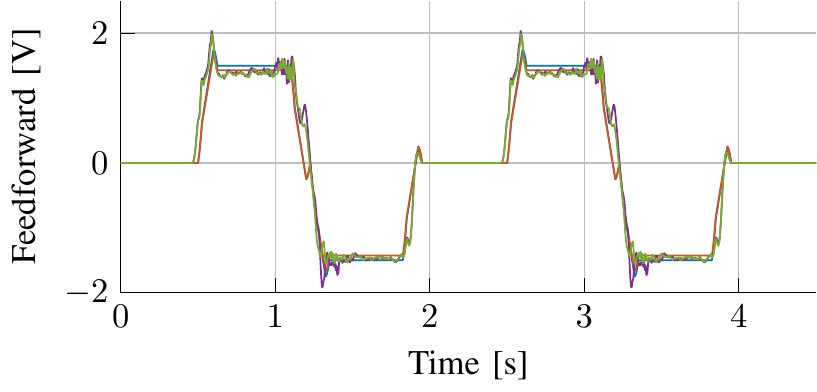}
\caption{Feedforward signals for ${r}_1$ produced by $F(q)$ (\protect\redline) and by the GP-based method (\protect\greenline), and for ${r}_2$ by $F(q)$ (\protect\blueline) and by the GP-based method (\protect\purpleline). The GP leads to a more complex signal, possibly to compensate for position-dependent static friction.}
\label{fig:ff}
\end{figure}

\subsection{Results}
The GP-based feedforward signals are shown in Figure \ref{fig:ff}. It can be seen that the GP-based feedforward signals exhibit more complex variations than the signal resulting from $F(q)$, which may indicate that nonlinear dynamics such as position-dependent friction are learned. Moreover, the GP-based feedforward signal acts earlier to a changing reference than $F(q)$, since it has $n_{ac}=40$ samples preview, whereas $F(q)$ has no preview. 

\begin{figure}[!t]
\centering
\includetikz{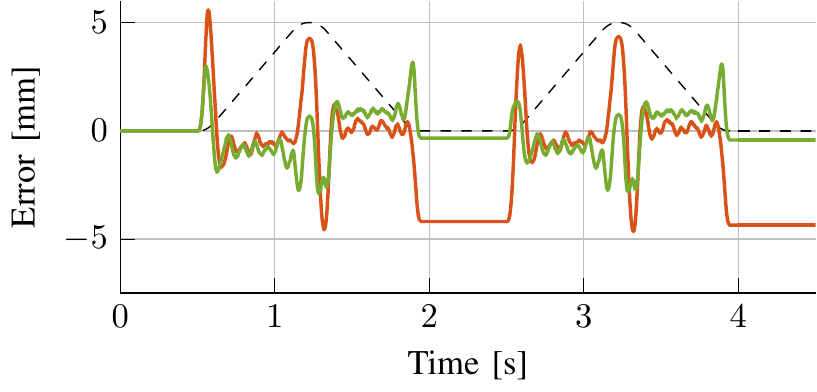}
\caption{Error ${e}_1={r}_1-{y}_1$ using $F(q)$ (\protect\redline) and the GP-based feedforward signal (\protect\greenline), along with the scaled reference (\protect\blackdash).}
\label{fig:error_nonflex}
\end{figure}

The feedforward signal is applied in closed-loop and the resulting error for ${r}_1$ is shown in Figure \ref{fig:error_nonflex}. The results are summarized in Table \ref{tab:printer}. It can be readily seen that the tracking error is improved significantly. In particular, the linear feedforward controller $F(q)$ leads to a large tracking error around $t=2$ s since it stops at an incorrect position, as a result of static friction. The nonlinear GP-based feedforward controller $\hat{F}(q)$, on the other hand, appears to have learned these dynamics, since the error at this point in time is reduced by a factor 12. 

\begin{figure}[!t]
\centering
\includetikz{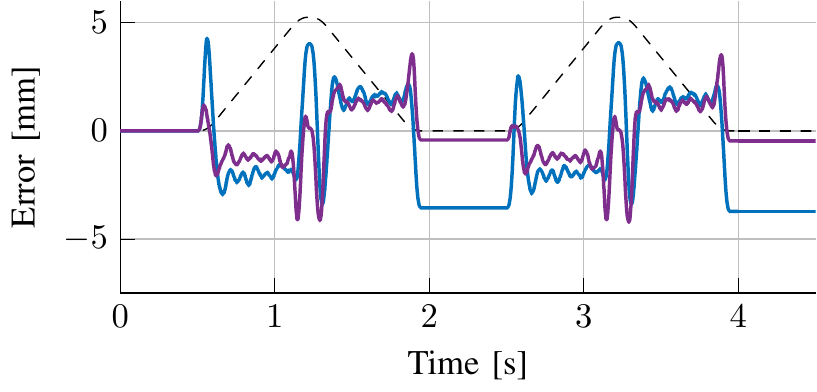}
\caption{Error ${e}_2={r}_2-{y}_2$ using $F(q)$ (\protect\blueline) and the GP-based feedforward signal (\protect\purpleline), along with the scaled reference (\protect\blackdash). While ${r}_2$ was not used as a reference when obtaining the dataset, the error is still reduced by the GP-based approach.}
\label{fig:error_flex}
\end{figure}

To demonstrate the ability of the method to deal with task flexibility, Figure \ref{fig:error_flex} shows the achieved performance for a reference $r_2$ not used during training. The tracking error is reduced to a similar extent as with $r_1$. This indicates that the method allows for the generation of feedforward signals for flexible tasks, provided that these tasks (or references) are sufficiently similar to the observed outputs in $\mathcal{D}$. This condition appears to be satisfied, as can be seen from Figure \ref{fig:ref}.

\begin{table}[]
\centering
\caption{Tracking errors with different feedforward signals.}\label{tab:printer}
\begin{tabular}{@{}llll@{}}
\toprule
                        &              & With ${r}_1\in\tilde{\mathcal{R}}$ & With ${r}_2\notin\tilde{\mathcal{R}}$ \\ \midrule
$\|{e}\|_2$      & $F(q)$       & 170 {[}mm{]}                                        & 164 {[}mm{]}                                                    \\
                        & $\hat{F}(q)$ & 68 {[}mm{]}                                         & 86 {[}mm{]}                                                     \\ \midrule
$\|{e}\|_\infty$ & $F(q)$       & 5.6 {[}mm{]}                                        & 4.2 {[}mm{]}                                                    \\
                        & $\hat{F}(q)$ & 3.2 {[}mm{]}                                        & 3.6 {[}mm{]}                                                    \\ \bottomrule
\end{tabular}
\end{table}


\section{Conclusions and future work}\label{sec:conclusions}
A method for inverse model control using Gaussian Process regression is developed that enables the generation of feedforward signals in the presence of nonlinear dynamics of unknown structure. It is shown experimentally that the approach can lead to improved tracking performance with respect to linear feedforward. Since this method requires tasks to be similar to observed outputs in the dataset, it can be particularly useful in situations when many slightly different references need to be tracked. Moreover, since the developed approach makes use of the available controllers $C(q)$ and $F(q)$ to construct an improved $\hat{F}(q)$, it can be seen as an add-on that is applicable without changing the control architecture.

Possible extensions include the application to multi-input multi-output systems, and a formal means to take output noise into account.

\addtolength{\textheight}{-12cm}   






\bibliographystyle{plain}
\bibliography{references}

\end{document}